\documentclass[12pt]{article}
\normalbaselines
\newcommand{\be}{\begin{equation}}
\newcommand{\ee}{\end{equation}}
\newcommand{\ba}{\begin{array}}
\newcommand{\ea}{\end{array}}
\newcommand{\ben}{\begin{enumerate}}
\newcommand{\een}{\end{enumerate}}
\newcommand{\bec}{\begin{center}}
\newcommand{\eec}{\end{center}}

\catcode `\@=11
\@addtoreset{equation}{section}

\def\section{\@startsection {section}{1}{\z@}{-3.5ex plus -1ex minus
     -.2ex}{2.3ex plus .2ex}{\normalsize\bf}}
\def\subsection{\@startsection{subsection}{2}{\z@}{-3.25ex plus -1ex minus
 -.2ex}{1.5ex plus .2ex}{\normalsize\bf}}
\def\@cite#1#2{${}^{\mbox{\scriptsize#1\if@tempswa , #2\fi}}$}
\def\thebibliography#1{\section*{References\markboth
  {REFERENCES}{REFERENCES}}\list
  {\arabic{enumi}.}{\settowidth\labelwidth{[#1]}\leftmargin\labelwidth
  \advance\leftmargin\labelsep
  \usecounter{enumi}}
  \def\newblock{\hskip .11em plus .33em minus -.07em}
  \sloppy
  \sfcode`\.=1000\relax}
 
\catcode `\@=12

\begin{document}
\noindent
{\bf MASSIVE RELATIVISTIC SYSTEMS WITH SPIN AND THE TWO TWISTOR PHASE SPACE.
\footnote{published in the proceedings of the conference HADRONS 96', Ukraine,
Crimea, Novy Svet, July 1996}}
\vskip 40pt
\noindent
\hspace*{1in}
\begin{minipage}{13cm} 
\bf Andreas Bette, \rm \\
Department of Physics, \\
Stockholm University, \\
Box 6730, \\
S-113 85 Stockholm, Sweden.

\vskip 20pt

\noindent
\bf Stanis{\l}aw Zakrzewski, \rm \\
Department of Mathematical Methods in Physics, \\
Warsaw University, \\
Ho{\.z}a 74, \\
Pl-00 682 Warsaw, Poland.
\end{minipage}

\vspace*{0.5cm}

\begin{abstract}
\noindent
We show that a two twistor phase space {\`a} priori describing two non
localized massless and spinning particles may be decomposed into a
product of three independent phase spaces: the (forward) cotangent
bundle of the Minkowski space, the cotangent bundle to a circle (electric
charge phase space) and the cotangent bundle to the real projective
spinor space. Reduction of this 16 dimensional phase space with
respect to two mutually commuting conformal scalars (the electric
charge and the difference between the two helicities) produces a 12
dimensional extended relativistic phase space\cite{zak}
describing a massive spinning particle.
\end{abstract}
 
\section{\hspace{-4mm}\hspace{2mm}TWISTOR PHASE SPACES.}

The fundamental object in twistor theory is the {\em twistor
space}\cite{jmp,int,oxf,pen-cal} $T$, which is a 4-dimensional complex
vector space equipped with a hermitian form:

\begin{equation} 
\rho:=Z^{\alpha}\bar W_{\alpha},
\end{equation}

\vskip 10pt

\noindent
where $Z,  W \ \epsilon \ T$ and where linear transformations of $T$
preserving this hermitian form and having the determinant equal to one
constitute a group $G$ isomorphic to $SU(2,2)$.  The important fact is that
the latter contains the (universal covering group of the) Poincar{\'e}
group as a subgroup.

\vskip 10pt

\noindent
The set of isotropic (with respect to the hermitian form)
two-dimensional subspaces in $T$ which are transversal to one
distinguished such a subspace defines the usual Minkowski space.

\vskip 10pt

\noindent
Relative to an arbitrary but fixed origin in the Minkowski space each
twistor may be represented by two spinors\cite{jmp,pen-cal}:

\begin{equation}
Z^{\alpha} = (\omega^{A},\ \pi_{A^\prime}). 
\end{equation}

\vskip 10pt

\noindent
Similarily the corresponding complex conjugate twistor is represented
by:

\begin{equation}
\bar Z_{\alpha} = ( \bar \pi_{A},\ \bar \omega^{A^\prime}).
\end{equation}

\vskip 10pt

\noindent
(Note that the splitting of a twistor into two spinors is translation
dependent i.e. depends on our choice of the fixed origin in the
Minkowski space.)

\vskip 10pt

\noindent
The SU(2,2) group is isomorphic with (the universal covering of) the
conformal group of the Minkowski space. Therefore the twistor space
provides a natural framework for the study of the conformal geometry
of the Minkowski space. Moreover, it has a natural structure of a {\em
phase space} of a massless particle with an arbitrary helicity
(spin). This follows from the fact that the imaginary part of the
hermitian form defines a $G$-invariant constant symplectic 
two-form\cite{magic} $\Omega$ on $T$:

\begin{equation}
\Omega = idZ^{\alpha}\wedge d\bar Z_{\alpha}.
\end{equation}

\vskip 10pt

\noindent
The canonical conformally invariant Poisson bracket algebra implied by
$\Omega$ gives the following commutation relations:

\begin{equation}\{{\bar  Z}_{\beta}, \ Z^{\alpha}\} 
= i{\delta}^{\alpha}_{\beta},
\end{equation} 

\vskip 10pt

\noindent
with all the remaining commution relations being equal to zero.  

\vskip 10pt

\noindent
The {\em two-twistor phase space} \bf Tp(2) \rm
is now defined\cite{ab} as the set of pairs of twistors:

\begin{equation}
Z^{\alpha} = (\omega^{A},\ \pi_{A^\prime}) \qquad and \qquad
W^{\alpha} = (\lambda^{A},\ \eta_{A^\prime})
\end{equation}

\vskip 10pt

\noindent
such that
\be 
f: = \pi^{A^\prime}\eta_{A^\prime} \neq 0.
\ee

\vskip 10pt

\noindent
The corresponding complex conjugate twistors are:

\begin{equation}
\bar Z_{\alpha} = ( \bar \pi_{A},\ \bar \omega^{A^\prime})
\qquad and \qquad 
\bar W_{\alpha} = (\bar \eta_{A},\ \bar \lambda^{A^\prime}).
\end{equation}

\vskip 10pt

\noindent
The conformally invariant symplectic structure $\Omega_{0}$ on \bf
Tp(2) \rm is now given by\cite{ab,tod,tod1}:

\begin{equation}
\Omega_{0} = 
i (dZ^{\alpha}\wedge d\bar Z_{\alpha} + dW^{\alpha}\wedge d\bar W_{\alpha}).
\end{equation}

\vskip 10pt

\noindent
The canonical conformally invariant Poisson bracket algebra implied by
$\Omega_{0}$ gives the following commutation relations:

\begin{equation}
\{{\bar  Z}_{\beta}, \ Z^{\alpha}\} 
= i{\delta}^{\alpha}_{\beta},
\qquad \qquad \qquad 
\{{\bar  W}_{\beta}, \ W^{\alpha}\} 
= i{\delta}^{\alpha}_{\beta},
\end{equation} 

\vskip 10pt

\noindent
with all the remaining commution relations being equal to zero.  In
terms of the Poincar{\'e} covariant spinor coordinates the only
non-vanishing commutations relations are:

\begin{equation}
\{{\bar  \pi_{B}}, \ \omega^{A} \} = i{\delta}^{A}_{B},
\qquad \qquad \qquad 
\{{\bar  \eta_{B}}, \ \lambda^{A} \} = i{\delta}^{A}_{B}.
\end{equation}

\vskip 10pt

\noindent
On \bf Tp(2) \rm there exist six independent real valued Poincar{\'e}
scalar functions\cite{ab,tod,tod1}. Four of these are also conformally
scalar.  The first two Poincar{\'e} (but not conformally)
scalar functions are those defined by $f$ in (1.7).

\vskip 10pt

\noindent
The four real valued conformally scalar functions are defined by means
of the the hermitian form introduced in (1.1):

\begin{equation}
s_{1} := {1 \over 2}(Z^{\alpha}\bar Z_{\alpha}) \qquad and \qquad 
s_{2} := {1 \over 2}(W^{\alpha}\bar W_{\alpha}),
\end{equation}

\begin{equation}
{\rho} := Z^{\alpha}\bar W_{\alpha}. 
\end{equation}

\vskip 10pt

\noindent
On each of the two single twistor phase spaces the four Lorentz
covariant and four translation invariant functions representing the
two linear null four momenta will be denoted by (abstract index
notation\cite{batelle,prr}):

\begin{equation}
P_{1a} := \pi_{A^\prime}{\bar\pi}_A  \ \ \ \
P_{2a} := \eta_{A^\prime}{\bar\eta}_A
\end{equation}

\vskip 10pt

\noindent
while Poincar{\'e} covariant functions (six for each single twistor
phase space) representing the two angular null four momenta will be
denoted by:

\begin{equation}
M_{1ab}:=i{\bar \omega}_{(A^\prime}\pi_{B^\prime )}\epsilon_{AB} + c.c.
\ \ \ \
M_{2ab}:=i{\bar \lambda}_{(A^\prime}\eta_{B^\prime )}\epsilon_{AB}+ c.c.
\end{equation}

\vskip 10pt

\noindent
(The above functions arise as generators of the Poincar{\'e} group
action, see below.)

\vskip 10pt

\noindent
By forming the Pauli Luba{\'n}ski spin four-vectors on each of the two
single twistor phase spaces one discovers that the two real valued
functions in (1.12) represent\cite{pen-cal} the classical limit of the two
helicity operators corresponding to each of the two massless systems
described by the two pairs $(P_{1}, \ M_{1})$ and $(P_{2}, \ M_{2})$.

\vskip 10pt

\noindent
In the two twistor phase space, by linearity, the above single twistor
phase space functions representing $(P_{1}, \ M_{1})$ and $(P_{2}, \
M_{2})$ define four new functions now representing a massive four
momentum and six new functions now representing a massive angular four
momentum\cite{ab,hugh}:

\begin{equation}
P_{a} := P_{1a} + P_{2a} 
\end{equation}

\begin{equation}
M_{ab}:= M_{1ab} + M_{2ab}.
\end{equation}

\vskip 10pt

\noindent
The absolute value of $f$ divided by the square root of two may now be
recognized as the rest mass of a system having its four-momentum
given by $P_{a}$.

\vskip 10pt

\noindent
It is a well established fact that the pair $(P_{1}, M_{1})$ and the
pair $(P_{2}, M_{2})$ each define separately a momentum mapping of the
Poincar{\'e} group into the corresponding single twistor phase space.
In other words, Poisson brackets (with respect to the conformally
invariant symplectic structure on a single twistor phase space) among
the functions defining each such a pair reproduce the algebra of the
Poincar{\'e} group\cite{oxf,ab,tod,hugh,ab1}. This automatically
implies that on the two twistor phase space the same is automatically
valid for the pair $(P, M)$.

\vskip 10pt

\noindent
On the two twistor phase space we may also distinguish the following
four mutually commuting complex valued functions representing the four
vector position coordinates in the complexified Minkowski space:

\begin{equation}
z^{a} := {i\over f}(\omega^{A}\eta^{A^\prime}-\lambda^{A}\pi^{A^\prime}).
\end{equation}

\vskip 10pt

\noindent
For future use we introduce the following mutually commuting functions
in \bf Tp(2) \rm representing three mutually Lorentz orthogonal space
like four-vectors in the Minkowski space:

\begin{equation}
{\l}_{a} := P_{1a} - P_{2a} 
\end{equation}

\begin{equation}
w_{a} := \pi_{A^\prime}{\bar\eta}_A.
\end{equation}

\vskip 10pt

\noindent
It is easy to see that the so defined four-vectors are also orthogonal
to the massive four momentum $P_{a}$.  Their Lorentz norm is equal to
minus the square of the rest mass of the system.  In addition the
functions which represent these three space like four-vectors commute
with the functions representing the massive four momentum vector
$P_{a}$. 

\vskip 10pt

\noindent
Forming the Pauli-Luba{\'n}ski spin four-vector on the two twistor
phase space:

\begin{equation}
S^{a}:={1 \over 2}\epsilon^{abcd}M_{bc}P_{d}
\end{equation}

\vskip 10pt

\noindent
yields\cite{ab,tod,hugh,ab1,perjes,sparling}:

\begin{equation} 
S_{a} = k{\l}_{a} +  \rho {\bar w}_{a} + {\bar \rho}w_{a},
\end{equation}

\vskip 10pt

\noindent
where 

\begin{equation}
k:=s_{1}-s_{2}.
\end{equation}

\vskip 10pt

\noindent
Reexpressing the massive angular four momentum as a sum of the spin 
and the orbital four angular momentum gives:

\begin{equation} 
M^{ab}=P^{a}X^{b} - P^{b}X^{a} + {1 \over (P^{k}P_{k})^{2}}\epsilon^{abcd}P_{c}S_{d},
\end{equation}

\vskip 10pt

\noindent
where $X$ is the real part of $z$:

\begin{equation}
X^{a} := {1\over 2}(z^{a} + {\bar z}^{a}).
\end{equation}

\vskip 10pt

\noindent
The imaginary part of $z$ may now be expressed as a Lorentz four-vector 
given by\cite{ab,tod,tod1,ab1}:

\begin{equation}
Y^{a} : = {1\over 2i}(z^{a} - {\bar z}^{a})
={1\over 2f\bar f}
(
\rho{\bar w}^{a} + 
{\bar \rho}w^{a}
- 
s_{1}P^{a}_{2}
- 
s_{2}P^{a}_{1}
).
\end{equation}

\vskip 10pt

\noindent
Note that the scalar spin function, defined in the usual way:

\begin{equation}
s:={1 \over m}{\sqrt {-S_{a}S^{a}}},
\end{equation}

\vskip 10pt

\noindent
is, as easily follows from (1.22), in terms of the twistor scalars
given by:

\begin{equation}
s = \sqrt {k^{2} + {\vert \rho \vert}^{2}},
\end{equation}

\vskip 10pt

\noindent
which shows that the spin of a massive particle is not only a Poincar{\'e}
but also a conformal scalar, the fact first time noticed by
Perj{\`e}s\cite{perjes}.  For massive particles formed by means of
more than two twistors this is no longer true\cite{ab,sparling}.

\vskip 10pt

\noindent
Besides the already mentioned Poisson bracket relations reproducing the
Poincar{\'e} algebra:

\begin{equation}\{P_{a},\ P_{b}\}=0,\end{equation}

\begin{equation}
\{M^{ab} ,\ P_{c}\}=
-P^{a}{\delta}^{b}_{c} 
+ P^{b}{\delta}^{a}_{c},
\end{equation}

\begin{equation}
\{M_{ab}, \  M_{cd}\}=
M_{ac}g_{bd} + M_{bd}g_{ac}
-M_{ad}g_{bc} - M_{bc}g_{ad},
\end{equation}

\vskip 10pt

\noindent
the canonical commutation relations in (1.10) and/or (1.11)
also imply that the functions representing the four position $X^{a}$
and the four momentum $P_{a}$ obey, as they should, the following
commutation relations:

\begin{equation}\{P_{a},\ X^{b}\}=\delta_{a}^{b}.\end{equation}

\vskip 10pt

\noindent
However, such four position functions allowing the splitting of the
four angular momentum into its orbital part and its pure spin part do
not, in general, commute, we have instead\cite{ab,zab}:

\begin{equation}
\{X^{a},\ X^{b}\}=-{1 \over (P_{k}P^{k})^{2}}R^{ab}
\end{equation}

\vskip 10pt

\noindent
where we have put:

\begin{equation}
R^{ab}:={1 \over (P_{k}P^{k})^{2}}\epsilon^{abcd}P_{c}S_{d}.
\end{equation}

\vskip 10pt

\noindent
This completes our short review of known facts following from
twistor theory. We now proceed to give a presentation of some
new facts. However, to make our exposition more transparent we do
not present any proofs. These may be found in another paper\cite{zab}.

\section{\hspace{-4mm}\hspace{2mm}THE COMMUTING FOUR POSITION FUNCTIONS.}

Is it possible to find four-vector valued function $\Delta X^{a}$ on
the two twistor space so that the functions representing the
shifted new four position defined by:

\begin{equation}
{\tilde X}^{a}:= X^{a} + \Delta X^{a}
\end{equation}

\vskip 10pt

\noindent
do commute, i.e. so that we obtain:

\begin{equation}
\{{\tilde X}^{a},\ {\tilde X}^{b}\}=0?
\end{equation}

\vskip 10pt

\noindent
Our first new result is that there exist two such shifts $\Delta
X^{a}$ (how these shifts were obtained will be described in another
paper\cite{zab}):

\begin{equation}
\Delta X^{a}:=\pm {i \over m^{2}}(\rho {\bar w}^{a} - {\bar \rho}w^{a}).
\end{equation}

\vskip 10pt

\noindent
By taking any of these two shifts the four angular momentum functions
in (1.24) may be rewritten as:

\begin{equation}
M^{ab}=P^{a}X^{b} - P^{b}X^{a} + R^{ab}
=P^{a}{\tilde X}^{b} - P^{b}{\tilde X}^{a}  + R^{ab} - V^{ab}
=P^{a}{\tilde X}^{b} - P^{b}{\tilde X}^{a}  + \Sigma^{ab},
\end{equation}

\vskip 10pt

\noindent
where

\begin{equation}
V^{ab}:=P^{a} \Delta X^{b} - P^{b} \Delta X^{a} 
\end{equation}

\vskip 10pt

\noindent
represents a sort of an "internal" orbital angular momentum relative to the
non-commuting four position, and where:

\begin{equation}
\Sigma^{ab}:=R^{ab} - V^{ab}.
\end{equation}

\vskip 10pt

\noindent
It is easy to deduce that:

\begin{equation}
\{{\tilde X}^{a},\ \Sigma_{bc}\}=0,
\end{equation}

\begin{equation}
\{P_{a},\ \Sigma_{bc}\}=0.
\end{equation}

\section{\hspace{-4mm}\hspace{2mm}THE DECOMPOSITION OF THE TWO TWISTOR PHASE SPACE.}

In this final section we show (without any explicit proofs which are
presented elsewhere\cite{zab}) how the two twistor phase space
decomposes in a way described in the abstract.

\vskip 10pt

\noindent
First we choose one of the two options in (2.3) e.g.:

\begin{equation}
\Delta X^{a}_{+}:={i \over m^{2}}(\rho {\bar w}^{a} - {\bar \rho}w^{a}).
\end{equation}

\vskip 10pt

\noindent
(Choosing the shift with the minus sign in (2.3) simply interchanges
the role of the two spinor variables $\eta$ and $\pi$.)

\vskip 10pt

\noindent
Then, by standard procedures\cite{prr}, it may be shown that
$\Sigma_{ab}$ is, in terms of spinors, given by:

\begin{equation}
\Sigma_{ab}:={\sigma}_{(A^\prime}\eta_{B^\prime )}\epsilon_{AB} + c.c.,
\end{equation}

\vskip 10pt

\noindent
where 

\begin{equation}
{\sigma}_{A^\prime}:=-{i \over f}(k \pi_{A^\prime}+\rho \eta_{A^\prime}).
\end{equation}

\vskip 10pt

\noindent
Introduce now two arbitrary but constant spinors $\alpha$ and $\beta$ such that

\begin{equation}
{\alpha}^{A^\prime}{\beta}_{A^\prime}=1,
\end{equation}

\vskip 10pt

\noindent
and consider following six real valued functions on \bf Tp(2) \rm:

\begin{equation}
k, \ \ \varphi \qquad and \qquad u_{1}, \ \ u_{2} 
\qquad and \qquad v_{1},  \ \  v_{2} 
\end{equation}

\vskip 10pt

\noindent
where 

\begin{equation}
e^{2i\varphi}:= 
{
\alpha^{B^\prime} 
\eta_{B^\prime} 
\over  {\bar \alpha}^{C}{\bar \eta}_{C}
},
\end{equation}

\begin{equation}
v:=v_1 + i v_2= 
{
\beta^{B^\prime} 
\eta_{B^\prime} 
\over \alpha^{C^\prime} 
\eta_{C^\prime} 
},
\end{equation}

\begin{equation}
u:=u_1 + i u_2= - 2
{(\alpha^{B^\prime} 
\sigma_{B^\prime}) 
(\alpha^{C^\prime} \eta_{C^\prime} )}.
\end{equation}

\vskip 10pt

\noindent
The above introduced functions are well defined on \bf Tp(2) \rm
except for the values of the variable spinor $\eta_{A^\prime}$ which
are proportional to the fixed values of the spinor
$\alpha_{A^\prime}$. If so happens one simply interchanges the role of
$\alpha$ and $\beta$.

\vskip 10pt

\noindent
Now it may be checked by direct computations that the functions
representing $\tilde X_{+}$, $P$, $s_{1}$ and $\arg f$ commute with
the functions $\varphi$, $k$, $u$ and $v$ while the only
non-vanishing Poisson brackets of the latter are:

\begin{equation}
\{k, \ \varphi\}=1, 
\qquad \qquad 
\{u_1, \ v_1 \}=1, 
\qquad \qquad 
\{v_2, \ u_2 \}=1,
\end{equation}

\vskip 10pt

\noindent
which shows that $k$, $u_1$, $v_2$ are canonically conjugate functions
to the functions $\varphi$, $v_1$, $u_2$.

\vskip 10pt

\noindent
We note, in passing,  that the functions representing $\Sigma$ are completely
defined by the functions $u$, $v$ and $k$.

\vskip 10pt

\noindent
The remaining non-vanishing Poisson brackets on \bf Tp(2) \rm are:

\begin{equation}
\{e, \ \phi \}=1,
\end{equation}

\vskip 10pt

\noindent
and 

\begin{equation}
\{P_{a},\ {\tilde X}^{b}\}=\delta_{a}^{b},
\end{equation}

\vskip 10pt

\noindent
where we have introduced:

\begin{equation}
e:=2s_{1} \qquad and  \qquad \phi:=\arg f.
\end{equation}

\vskip 10pt

\noindent
Expressed in terms of these new coordinates 
the symplectic structure in (1.9) reads:

\begin{equation}
\Omega_{0} = 
dP_{a}\wedge d{\tilde X}^{a}_{+} + de \wedge d\phi +
dk \wedge d\varphi+ du_{1} \wedge dv_{1}+ dv_{2} \wedge du_{2}.
\end{equation}

\vskip 10pt

\noindent
In effect we have three sets of mutually commuting variables each set
defining its own symplectic manifold. The first has eight dimensions
and is spanned by ${\tilde X}_{+}$ and $P$, the second two dimensional
is spanned by $e$ and $\phi$ and the third six dimensional is defined
by $k$, $u_1$, $v_2$ and $\varphi$, $v_1$, $u_2$.  Reduction with
respect to the conformally scalar variables $k$ and $e$ gives a twelve
dimensional phase space identical\cite{zab} with the extended phase
space (the case with $b=0$) introduced in Ref. 1.

\end{document}